\documentclass{PoS}
\usepackage{graphicx}
\usepackage[authoryear,round]{natbib}

\title{The SKA Mid-frequency All-sky Continuum Survey: Discovering the unexpected
and transforming radio-astronomy}

\ShortTitle{SKA All-sky survey}
\author{
\speaker{Ray P.\ Norris}$^{1,7}$,
Kaustuv Basu$^2$,
Michael Brown$^3$,
Ettore Carretti$^1$,
Anna D. Kapi\'{n}ska$^{4,7}$,
Isabella Prandoni$^5$,
Lawrence Rudnick$^6$,
Nick Seymour$^{1,4}$
 \\
$^1$CSIRO Astronomy \& Space Science, PO Box 76, Epping, NSW 1710, Australia;
$^2$Argelander Institute for Astronomy, University of Bonn, Auf dem Huegel 71, 53121 Bonn, Germany;
$^3$Michael J. I. Brown, School of Physics, Monash University, Clayton Vic 3800, Australia;
$^4$ICRAR, The University of Western Australia, 35 Stirling Hwy, Crawley, WA 6009, Australia;
$^5$ INAF - Istituto di Radioastronomia, Via Gobetti 101, 40125 Bologna, Italy;
$^6$Minnesota Institute for Astrophysics, School of Physics \& Astronomy, University of Minnesota, USA.
$^7$ARC Centre of Excellence for All-Sky Astrophysics (CAASTRO), Sydney, Australia.\\
E-mail: \email{Ray.Norris@csiro.au}

}

\abstract{{\bf SKA is an instrument, not an experiment} (Phil Diamond, Stellenbosch, 17 Feb 2014)\\
 
We show that, in addition to  specific science goals, there is a strong case for conducting an all-sky  (i.e.  the visible $3\pi $ steradians) SKA continuum survey which does not fit neatly into conventional science cases.
History shows that the greatest scientific impact of most major telescopes (e.g., HST, VLA) lies beyond the original goals used to justify the telescope. The design of the telescope therefore needs  to maximise the ultimate scientific productivity, in addition to achieving the specific science goals. 
 In this chapter, we show that an all-sky continuum survey is likely to achieve transformational science in two specific respects:
\begin{itemize}
\item Discovering the unexpected
\item Transforming radio-astronomy from niche to mainstream
\end{itemize}}

\FullConference{
Advancing Astrophysics with the Square Kilometre Array\\
June 8-13, 2014\\
Giardini Naxos, Sicily, Italy}

\newcommand{\skipthis}[1]{}

\newcommand\apj{ApJ}
\newcommand{\kms}{\mbox{km\,s$^{-1}$}}

\def\kms {\ifmmode{{\rm ~km~s}^{-1}}\else{~km~s$^{-1}$}\fi}

\def\lsun {\ifmmode{{\rm ~L}_\odot}\else{~L$_\odot$}\fi}

\def\deg {^{\circ} }
\def\sqdeg {\,deg$^2$}

\def\ujybm {\,$\mu$Jy/beam}

\def\um {\,$\mu$m}

\newbox\grsign \setbox\grsign=\hbox{$>$} \newdimen\grdimen \grdimen=\ht\grsign
\newbox\simlessbox \newbox\simgreatbox
\setbox\simgreatbox=\hbox{\raise.5ex\hbox{$>$}\llap
 {\lower.5ex\hbox{$\sim$}}}\ht1=\grdimen\dp1=0pt
\setbox\simlessbox=\hbox{\raise.5ex\hbox{$<$}\llap
 {\lower.5ex\hbox{$\sim$}}}\ht2=\grdimen\dp2=0pt
\def\simgreat{\mathrel{\copy\simgreatbox}}

\def\lsim{\mathrel{\rlap{\lower4pt\hbox{\hskip1pt$\sim$}}
    \raise1pt\hbox{$<$}}}                
\def\gsim{\mathrel{\rlap{\lower4pt\hbox{\hskip1pt$\sim$}}
    \raise1pt\hbox{$>$}}}                

\def\apj {{\it Ap.~J.}}
\def\apjl {{\it Ap.~J.\ (Letters)}}
\def\apjs {{\it Ap.~J.\ Suppl.}}
\def\aj {{\it A.~J.}}

\def\aap {{\it Astr.~Ap.}}

\def\mnras {{\it MNRAS}}

\def\nat {{\it Nature}}
\def\pasa {{\it PASA}}

\def\aj{AJ}                   
\def\apj{ApJ}                 
\def\apjl{ApJ}                
\def\apjs{ApJS}               
\def\aap{A\&A}                
\def\mnras{MNRAS}             

\def\nat{Nature}              

\def\grtsim{\mathrel{\hbox{\rlap{\hbox{\lower2pt\hbox{$\sim$}}}\raise2pt\hbox{$>$}}}}
\def\lesssim{\mathrel{\hbox{\rlap{\hbox{\lower2pt\hbox{$\sim$}}}\raise2pt\hbox{$<$}}}}

\def\lsim{\,\lower2truept\hbox{${<\atop\hbox{\raise4truept\hbox{$\sim$}}}$}\,}
\def\gsim{\,\lower2truept\hbox{${> \atop\hbox{\raise4truept\hbox{$\sim$}}}$}\,}
\def\simlt{\mathrel{\rlap{\lower 3pt\hbox{$\sim$}}
        \raise 2.0pt\hbox{$<$}}}
\def\simgt{\mathrel{\rlap{\lower 3pt\hbox{$\sim$}}
        \raise 2.0pt\hbox{$>$}}}

\begin{document}

\section{Introduction }


Experience shows that when telescopes enter unexplored areas of observational phase space, they make unexpected discoveries, and these discoveries often outshine the specific goals for which the telescope was built. For example, only one of the top ten discoveries with the Hubble Space Telescope ({\it HST}) was listed amongst the key goals used to justify its funding. So while specific science goals are useful to focus SKA design, they are unlikely to appear amongst its greatest scientific achievements. So in addition to achieving {\bf known} science goals, SKA must be designed to maximise its ability to discover the (potentially more important){ \bf unknown} science goals. 
An all-sky survey maximises the chance of detecting rare unknown objects and phenomena.

An additional motivation for an all-sky survey is that radio-astronomy is currently something of a niche science, and major radio-astronomical surveys typically  have about 10\% of the impact of major optical and infrared surveys. This is set to change as next-generation radio surveys cross a threshold beyond which they are dominated by normal star-forming galaxies, as opposed to the much rarer radio-loud AGN which have dominated previous radio surveys.

Both these general goals, and other specific science goals, are best addressed by an all-sky radio continuum survey.
For convenience in this paper, we abbreviate this proposed SKA1 All-Sky continuum Survey to SASS1, and designate as SASS2 the  potential counterpart with SKA2.

\citet{White14} points out that large survey telescopes (e.g. {\it WISE, GALEX, Pan-STARRS, LSST}) go to extraordinary lengths to achieve near-all-sky coverage.  The {\it 2MASS} project even went to the trouble of building observatories in both hemispheres to achieve true all-sky coverage. SASS1 will open up the radio sky just as other large all-sky surveys are opening up the optical, infrared, and X-ray skies. SASS1 is an opportunity for SKA to join the multiwavelength all-sky astronomical renaissance which will occur in the next decade.

Here we show that SASS1 is likely to achieve transformational science in two specific respects:
\vspace{-7mm}
\begin{itemize}
\itemsep-7pt
\item Discovering the unexpected
\item Transforming radio-astronomy from niche to mainstream
\end{itemize}
In addition, we list the supporting science cases for which an all-sky survey is critical for achieving specific science goals:
\begin{itemize}
\itemsep-7pt
\item Cosmology (Dipole/low-l multipoles, low-l Planck anomalies, $f_{NL}$, etc.)
\item Galaxy Clusters and Large-scale Structure
\item Evolution of Galaxies
\item The Magnetic Sky
\item Nearby Galaxies
\end{itemize}

This paper is organised as follows: \S 2 describes the Science Background, \S 3 discusses the transformational science cases, and \S 4 discusses the supporting science cases. \S\S 2 to 4 are written in the context of the SKA1 baseline design. This is summarised in \S 5, and then \S\S 6 and 7  discuss the science in two alternative scenarios: a 50\% SKA1, and SKA2.

\vspace{-5mm}
\section{Science Background}
\subsection{Radio Continuum Surveys}

\begin{figure}[h]
\begin{center}
\includegraphics[width=8cm, angle=0]{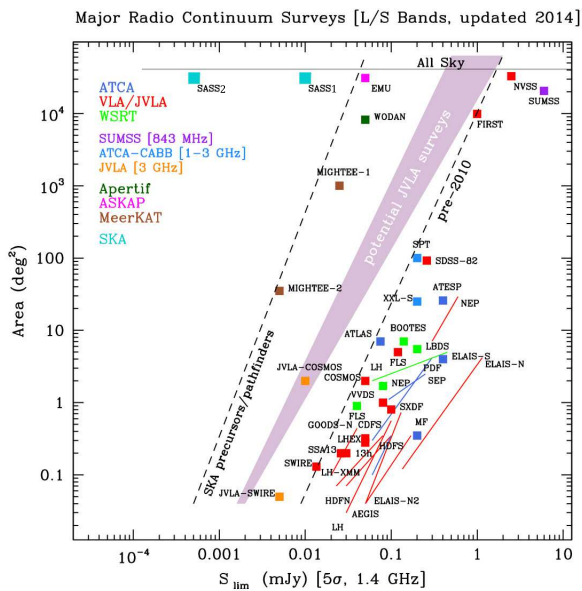}
\caption{Comparison of existing and planned deep 20 cm radio continuum surveys. The horizontal axis shows the  5-$\sigma$ sensitivity, and the vertical axis shows the sky coverage. The right-hand diagonal dashed line shows the approximate envelope of existing surveys, which is largely determined by the availability of telescope time.  
The squares in the top-left represent the new radio surveys discussed in this paper.  
}
\label{fig1}
\end{center}
\end{figure}

Radio continuum surveys have a rich and productive history, generating some of the most cited papers in radio-astronomy. Previous radio  surveys have  been dominated by AGN (Active Galactic Nuclei) but radio surveys are now crossing a threshold where normal star-forming galaxies dominate the number counts. For example, the EMU (Evolutionary Map of the Universe) survey on the Australian SKA Pathfinder (ASKAP) is expected to detect about 70 million galaxies (compared
to the current total of $\sim$ 2.5 million known radio sources), 
for most of which the radio emission is dominated by star formation rather than AGN.

Not only will next-generation radio surveys measure intensities to unprecedented levels, but they will also have better resolution, better sensitivity to extended emission, and will
measure spectral index and polarisation for the strongest 
sources. For example, the ASKAP surveys will measure polarisation and spectral index for about 3 million  sources, giving a 100-fold increase in the number of known polarised radio sources. 

The predicted sensitivities and areas for the main  1.4 GHz surveys are shown in Figure 1. The largest existing radio survey, shown in the top right, is the wide but shallow NRAO VLA Sky Survey \citep[NVSS:][]{Condon98}. The most sensitive existing radio survey is the deep but narrow JVLA-SWIRE (Lockman hole) observation in the lower left \citep{Condon12}. 
Current surveys are bounded by a diagonal line that roughly marks the limit of available telescope time of current-generation radio telescopes. The region to the left of this line is currently unexplored, and this area of observational phase space presumably contains as many potential new discoveries as the region to the right.

Wide-field survey radio astronomy in the next few years is likely to be dominated by ASKAP surveys, for which planning, funding, and construction is well advanced. For example, EMU \citep{Norris11} will survey 75\% of the sky to a sensitivity of 10\,\ujybm\ rms. Only a total of about 10\,\sqdeg\  of the sky has been surveyed at 1.4\,GHz to this sensitivity, in fields such as the {\it Hubble}, {\it Chandra}, ATLAS, COSMOS and Phoenix deep fields.  

From the early 2020's, SASS1 and other SKA1 radio surveys will advance well beyond the limits reached by current telescopes.  SASS1 will survey the sky to an rms of 2 \ujybm\ with a resolution of 2 arcsec. SASS1 will take 2 years with SKA1-SUR, or would take 6 years with SKA1-MID, or 50 years with ASKAP, or 600 years with the JVLA. 
We consider it unlikely that SKA1-MID would be scheduled for 6 years integration time for an all-sky survey, so for the rest of this document we use the SKA1-SUR specifications, for which we consider a 2-year time allocation to be realistic, especially if commensal with HI and polarisation surveys.

\subsection{The radio sky at $\mu$Jy levels}
Most extragalactic radio continuum surveys aim to understand the formation and evolution of galaxies over cosmic time, and the cosmological parameters and large-scale structures that drive it. 
Four generations of all-sky surveys are shown in Table \ref{surveytable}.

\begin{figure}
\begin{center}
\includegraphics[width=8cm, angle=0]{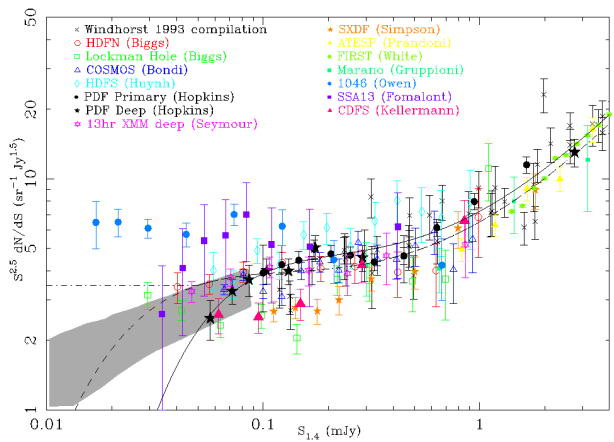}
\caption{The Euclidean normalised differential radio source counts at 1.4 GHz \citep[taken from][]{Norris13, Hopkins03}. The curves are alternative polynomial fits. The horizontal dot-dashed line represents a non-evolving population in a Euclidean universe. The shaded region shows the prediction based on fluctuations due to weak confusing sources \citep[$P(D)$ analysis:][]{Condon74, Mitchell85}.  }
\label{srccnt}
\end{center}
\end{figure}

\begin{table*}
\caption{Four generations of all-sky (i.e. $\sim 3 \pi$ steradian) continuum surveys. Counts for EMU are based on source counts from ATLAS \citep{Franzen, Banfield} and COSMOS \citep{Schinnerer}, and deeper counts are based on the $P(D)$ analysis by \citet{Vernstrom}. SKA1 specifications are taken from the baseline design (i.e. FOV 18 sq deg.). SKA2 specifications assume FOV=360 sq deg, and ten times the SKA1 sensitivity.
}
\begin{tabular}{lllllll}
\hline
Survey  & Resolution& 5$\sigma$ Flux & number &number of & Integration &reference\\
 &(arcsec)&density limit & of sources   &   pol. sources &time&\\
& & (\ujybm)& (millions)&(millions) &years & \\
\hline
NVSS &45	& 2250 & 1.8 & 0 & 0.31 & Condon et al.(1998)\\
EMU   &10  &      50     &      70     &         3 & 1.5 & Norris et al.(2011)\\
SASS1 & 2 &10    &    500     &      10 & 2 & this paper\\
SASS2  & 0.1         & 0.5    &      3500   &     70 & 0.5 & this paper\\
\hline
\end{tabular}
\label{surveytable}

\end{table*}

At high flux densities, the source counts (Figure 2) are dominated by AGN. Below 1 mJy/beam, the normalised source counts flatten, suggesting an additional population consisting of a mixture of both SF galaxies and radio-quiet AGN. It is difficult to distinguish AGN from SF galaxies and techniques include radio morphology, spectral index, polarisation, variability, radio--infrared ratio, optical and IR colours and spectral energy distributions (SED's), optical line ratios, X-ray power and hardness ratio, and radio brightness on VLBI scales. None of these techniques is foolproof, and a combination of techniques is necessary to provide unambiguous classification.

Furthermore, there is growing recognition that high-luminosity galaxies, particularly at high redshift, are not simply ``star-forming'' or ``AGN'' but include a significant contribution from both \citep[e.g.][]{Norris12}. Such galaxies are sometimes labelled ``composite'' galaxies. 

\begin{figure}
\begin{center}
\includegraphics[width=8cm, angle=0]{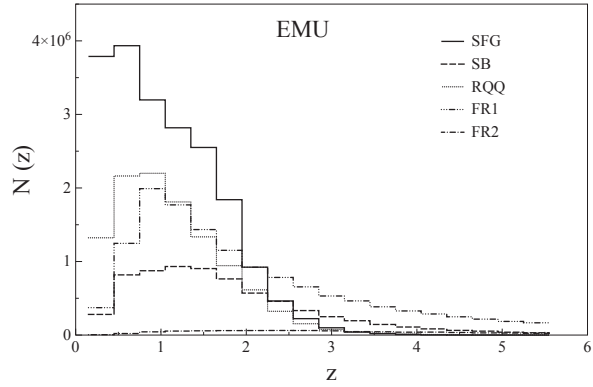}
\caption{Expected redshift distribution of  sources with $S_{1.4} > 50$\,\ujybm\ (the EMU $5-\sigma$ sensitivity), based on the SKADS simulations  \citep{Wilman08}. The five lines show the distributions for star-forming galaxies (SFG), starburst galaxies (SB), radio-quiet quasars (RQQ), and radio-loud galaxies of Fanaroff-Riley types I and II \citep{FRI}. 
}
\label{nz}
\end{center}
\end{figure}

\subsection{Cross-Identification}
\label{crossid}
Most science goals require cross-identification between radio sources and optical/infrared surveys. Surveys such as  EMU have relatively large synthesised beamwidths ($\sim$ 10 arcsec) which make this challenging, but this is mitigated by the following factors:
\vspace{-10pt}
\begin{itemize}
\itemsep-7pt
\item most radio sources at $\>50$\ujybm\ have  a 3.6 \um\ infrared counterpart, and so cross-identification between the radio and IR, and then cross-identification between IR and optical, produces a much better reliability than matching between radio and optical directly, and
\item the  accuracy of a radio source position is $\sim beamwidth/(2\times SNR)$, where SNR is the signal-to-noise ratio of the detection. So a 5-$\sigma$ detection has a positional accuracy of $\sim$ 1 arcsec, and stronger sources even better accuracy, and
\item using sophisticated Bayesian algorithms \citep[e.g.][]{Fan, Weston} which make full use of available photometric and morphological information, rather than using a simple nearest-neighbour  or likelihood ratio algorithm.
\end{itemize}
\vspace{-2mm}
As a result, cross-identifications with EMU are expected to achieve a high reliability.

For SASS1, the increased radio sensitivity  implies that detected galaxies will be (a)  optically fainter and (b) much more numerous. A higher spatial resolution is therefore needed, and the 2-arcsec synthesised beamwidth of SASS1 provides a 0.2 arcsec positional accuracy for a 5-$\sigma$ detection, ensuring  a high reliability of cross-identification.


\subsection{Redshifts}
Spectroscopic redshifts are currently known for about 100,000 extragalactic radio sources, and this will increase to about 1 million within the next five years. Thus most known radio sources do not have spectroscopic redshifts. However, many more non-radio galaxies (currently over 2 million, and expanding to tens of millions within a decade) have spectroscopic redshifts. As the sensitivity of radio surveys increases, the  fraction of radio sources that have measured spectroscopic redshifts will increase, but will continue to be a small fraction.

About half the SASS1 sources will have good optical/IR identifications with multiwavelength data (e.g. {\it SDSS, VHS, SkyMapper}), enabling a photometric redshift to be determined. In some cases, with good data, photometric redshifts can be extremely reliable, whereas those obtained from poor data are less reliable.

The large numbers (hundreds of millions) of sources enable new approaches to be used, such as statistical redshifts, in which the redshift of an individual galaxy is poorly known, but the redshift distribution of a sample of galaxies can be used  within a statistical framework.

Different science goals have different needs for redshifts. The cosmological tests described below yield significant constraints on cosmological parameters with no redshift knowledge whatsoever \citep{Raccanelli14}, although even incomplete redshift information significantly improves the constraint \citep{Camera12}. For goals such as measuring the evolution of cosmic star formation rate, photometric or statistical redshifts are adequate, provided the incompleteness and uncertainties are well-determined.

\section{Transformational Science Cases}

\subsection{Discovering the unexpected}
\label{WTF}

\citet{Ekers09} has shown that of 18 major astronomical discoveries in the last 60 years, only seven were planned. The remaining 11 were unexpected discoveries resulting from new technology or from observing the sky in an innovative way, exploring uncharted parameter space. In particular, the greatest science impact of new astronomical facilities often come not from the science goals listed in the proposal to build the telescope, but from unexpected discoveries; unexpected discoveries by a new instrument often outshine its original science goals. For example, Table \ref{HST} shows that only one of the top ten discoveries made with the {\it HST} was listed in the key science goals used to justify the project. While specific SKA science goals are necessary to focus the design of a scientifically productive facility, we must also recognise that the most significant discoveries from the SKA are unlikely to come from these science goals.

\begin{table*}
\caption{Major discoveries made by the Hubble Space Telescope ({\it HST}). Of the {\it HST}'s ``top ten'' discoveries (as ranked by National Geographic magazine), only one was a key project used in the {\it HST} funding proposal \citep{Lallo}.  A further four projects were planned in advance by individual scientists but not listed as key projects in the {\it HST} proposal. Half the ``top ten'' {\it HST} discoveries were unplanned, including two of the three most cited discoveries, and including the only {\it HST} discovery (Dark Energy) to win a Nobel prize.}
\vspace{2mm}
\begin{tabular}{llllll}
\hline
Project  & Key  & Planned? & Nat Geo 
 & Highly  & Nobel \\
  & Project? & &top ten? & cited? & Prize? \\
\hline
Use cepheids to improve value of $H_{0} $&\checkmark  & \checkmark &\checkmark  & \checkmark & \\
UV spectroscopy of ig medium  &\checkmark  & \checkmark & & & \\
Medium-deep survey &\checkmark  & \checkmark & & & \\
Image quasar host galaxies & &\checkmark  & \checkmark & & \\
Measure SMBH masses & & \checkmark &\checkmark  & & \\
Exoplanet atmospheres & & \checkmark &\checkmark  & & \\
Planetary Nebulae & &\checkmark  &\checkmark  & & \\
Discover Dark Energy & & & \checkmark &\checkmark  &\checkmark  \\
Comet Shoemaker-Levy & & & \checkmark & & \\
Deep fields (HDF, HDFS, GOODS, FF, etc) & & & \checkmark &\checkmark  & \\
Proplyds in Orion  & & & \checkmark & & \\ 
GRB Hosts & & &\checkmark  & & \\

\hline
\end{tabular}
\label{HST}

\end{table*}

On the other hand, discovering the unexpected need not be a random or passive process. Recognising the importance of unexpected discoveries, the SKA should be designed to optimise the survey strategy and data-mining software, to maximise the probability of such discoveries. This is achieved by  maximising the volume of virgin observational phase space, and  developing software to mine the data for the unexpected. SKA1 will open up new swathes of observational phase space, with a high likelihood of making significant unexpected discoveries. Searching for the unexpected, and developing software to mine for the unexpected, must be a high priority of SKA1.

Figure 1 shows that SASS1  will open up a large area of virgin parameter
space, to the left of the the dotted line, and should therefore plan for unexpected discoveries. We cannot rely on blind serendipity, because of the large data volumes, and the complexity of the instrument, but should plan to  mine the data systematically for the unexpected. To maximise the scientific productivity,  the SKA design process should:
\vspace{-3mm}
\begin{itemize}
\itemsep-5pt
\item start by designing the telescope to address known science goals
\item choose design parameters, observational parameters, and survey parameters to maximise the volume of new observational phase space
\item ensure that processing techniques are not limited to answering known questions
\item design software specifically to mine the data for unexpected discoveries.
\end{itemize}
Discoveries are thinly distributed through the observational phase space.  We cannot predict where they lie, and it is difficult to quantify the volume of phase space being explored. Key elements to maximising the expected number of unexpected discoveries will include 
\vspace{-3mm}
\begin{itemize}
\itemsep-5pt
\item making an all-sky survey to discover rare objects, 
\item making a deep survey to discover faint objects, 
\item maximise the use of relatively unexplored parameters such as circular polarisation, time variability, sensitivity to diffuse emission, etc.
\end{itemize}
 The last three items are already well addressed in other chapters. Here we emphasise the need for an all-sky survey to discover  rare objects.

\subsection{Transforming radio-astronomy from niche to mainstream}

It is not widely appreciated by radio-astronomers that radio-astronomy is seen as a niche area of astronomy by the majority of non-radio astronomers, because existing radio measurements  are not as intrinsically deep as existing optical data, and so $\sim$ 90\% of objects studied by a typical astronomer have no radio data. As a result, astronomers modelling the SED of a galaxy, or fitting a photometric redshift, will typically use optical and infrared data, but will rarely use radio data, because previous all-sky radio surveys were insufficiently sensitive to detect most galaxies. 
Similarly, developers of algorithms and templates don't bother including radio data in their code, making it even harder  to use radio data. As a result, the most-cited radio surveys have only about 10\% of the citations (and presumably only 10\% of the impact) of the most-cited optical surveys (Table \ref{cites}).

Next-generation radio continuum surveys with the SKA and its pathfinders are crossing a sensitivity threshold below which most galaxies detected in radio surveys are normal star-forming galaxies. For a  star-forming galaxy in the SKADS simulation \citep{Wilman08}, the 10 \ujybm\ detection limit of SASS1 corresponds to a median $I$-band magnitude of $\sim$ 21.5. Assuming typical R-I values of 0.5-1 \citep[e.g.][]{Smail95}, this corresponds to an R limit of 22.5-23, which is approaching the sensitivity limit of sky surveys such as {\it SDSS} and {\it SkyMapper}.
 The all-sky continuum survey with SKA2 will go several magnitudes deeper, approaching the survey limit of {\it LSST}. As a result, almost every galaxy found in optical/IR surveys will have radio photometry (and in many cases, polarisation and spectral index information). The high spatial resolution of SASS1 also ensures reliable cross-identifications with optical/IR hosts (see \S\ref{crossid}).

In 10 years time radio photometry is likely to be as commonplace in galaxy SED and photo-z estimation as IR photometry is now, leading to (a) better photometric redshifts and (b) a better distinction between star-forming and AGN components of the SED. This in turn will help address  many other science goals such as the evolution of the cosmic star formation, assembly of galaxies, and the role of AGN in regulating galaxy formation.

The radio data will add information orthogonal to that currently available, since radio is unaffected by dust and is a sensitive tracer of both star formation and AGNs 
even in so-called radio-quiet AGNs. Surveying the entire sky ensures that most objects of  interest in next generation surveys such as {\it LSST} have measured radio photometry.

In addition, polarisation and spectral index can be measured for tens of millions of stronger sources, to determine galaxy properties. For example, the detection of significant polarisation in anything but a very nearby galaxy demonstrates the presence of an AGN (and can be used to probe its properties) while spectral index can be used as an indicator of age in both star-forming galaxies (measuring the ratio of synchrotron to free-free emission) and in an AGN (by measuring the turnover frequency and energy loss by high-energy electrons). 

Radio-astronomy is therefore set to 
take its place alongside optical and infrared as another tool in every astronomer's toolkit, and an indispensable part of every fitted spectral energy distribution or photometric redshift measurement. Next-generation radio surveys are expected to have as many citations (and as much scientific impact) as next-generation optical surveys. SASS1 will not only increase the number of known radio sources by nearly an order of magnitude, but will dominate international  surveys, and move radioastronomy from niche science  to mainstream.

\begin{table}
\caption{The six most highly cited optical/infrared extragalactic surveys and the  six most highly cited radio extragalactic surveys, showing the  number of refereed journal papers based on the survey, and the number of papers that cite the survey. These numbers were measured by searching ADS for refereed papers which contained (``galaxy" or ``galaxies") and (``long survey name'' or ``short survey name'') in the abstract or title. Numbers for 3CR, and 4C are probably overestimates because they include papers which mention a source name starting with 4C etc. FIRST is not included  because the  name ``FIRST" produced an  unmanageable number of spurious results, and the long name of the FIRST survey does not make it into the top six.
}
\vspace{0.2cm}
\begin{tabular}{lllllll}
\hline
Survey & \# publications & \# citations & |   & Survey & \# publications & \# citations\\
\hline
& {\bf Optical/IR} & & | & &  {\bf Radio} & \\
\hline
SDSS & 4487 & 198454 &  |  & 4C &  458 & 13831\\
HDF & 601 & 52486 & | & NVSS &  355 & 11891\\
2MASS & 1167 & 38129 & | & 3CR &  323 & 15231\\
GOODS & 608 & 36100 & |  & HIPASS &  92 & 3441\\
2dFGRS & 210 & 27364 & | & PMN &  90 & 2603\\
CfA & 343 & 20929 & | & WENSS  &  64 & 1580\\
\hline

\end{tabular}
\label{cites}
\end{table}


\section{Supporting Science Cases}
In this Section we briefly discuss the specific science goals for which an all-sky survey delivers well-defined advantages over a smaller area survey.
 \subsection{Cosmology}
 \label{cosmology}

 SASS1 will make sensitive measurements of dark energy evolution, modified gravity and primordial non-Gaussianity, using a combination of probes listed in Table \ref{cosmo}., placing significant constraints on cosmological (e.g. Dark Energy equation of state) and fundamental physical parameters (e.g. departures from General Relativity, and non-Gaussian inflation), even when redshifts for individual radio sources are unknown 
 \citep[e.g.][]{Raccanelli10, Camera12, Raccanelli14, Rees14}. 

\begin{table*}
\caption{The four  cosmological probes for which SASS1 is competitive. A high-z ISW detection would be inconsistent with standard $\Lambda$CDM but consistent with massive neutrinos.
}
\vspace{0.2cm}
\begin{tabular}{lll}
\hline
& Technique & Physical effect \\
\hline
1. & Auto-correlations of radio data & spatial power spectrum\\
2. &Cross-correlation between (z $\leq$ 0.5) optical  & cosmic magnification at low z\\
& foreground galaxies and (<z> $\sim$ 1.5) sources\\
3. & Cross-correlation between sources & cosmic magnification at high z \\
& and CMB ($ \theta \leq 1\deg$; no z needed)\\
4. & Cross-correlation between source density  & Integrated Sachs-Wolfe effect\\
 & and CMB ($ \theta \sim 10\deg$) in 2-3 z bins \\
\hline
\end{tabular}
\label{cosmo}

\end{table*}

 In such studies, an all-sky survey is essential to minimise cosmic variance and measure low-order spherical harmonics. For example, radio source counts have already been shown \citep{Blake02} to have a dipole signature, and SKA surveys will provide a measurement of  that dipole signature that is independent of the dominant signal in WMAP/Planck. Furthermore, the Planck survey has shown a possible large-scale (low-l multipole) deviation from the standard $\Lambda$CDM model which can be tested with radio source counts over a survey covering a large fraction of the sky.
 The large sky area of the survey and the huge number of galaxies that will be detected, combined with their high mean redshift, will test parts of parameter space that are not accessible to other wavelengths.

\begin{figure}
\begin{center}
\includegraphics[width=8cm, angle=0]{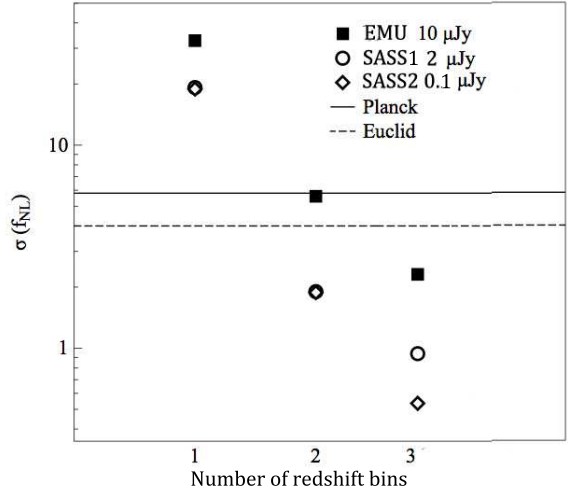}
\caption{The uncertainty of the measurement of the non-Gaussianity parameter $f_{NL} $ using the Integrated Sachs-Wolf effect on EMU, SASS1, and SASS2 data, compared to {\it Planck} and {\it Euclid}. Adapted from \citet{Raccanelli14}. }
\label{fnl}
\end{center}
\end{figure}

While some of these cosmological tests require no redshifts, even limited redshift information can dramatically improve them
\citep[e.g.][]{Raccanelli10, Camera12,  Raccanelli14, Rees14}. While obtaining individual redshifts for millions of radio sources is impossible in the next decade, dividing the radio source population into redshift bins is achievable with existing data. 
For example, selecting polarised radio sources without an optical identification in SDSS or Skymapper (i.e. $R>23$) yields a population of AGNs at z $\simgreat $1.  
This high redshift tail of the radio galaxy population can provide exquisite constraints on the evolution of dark energy. Separating the radio galaxy populations  and measuring the bias on ultra-large scales will give unprecedented constraints on primordial non-Gaussianity. 
 
\subsection{Galaxy Clusters and Large-Scale Structure}
\label{clusters}

Galaxy clusters are extremely sensitive probes of the growth rate of
cosmic structures, but finding and characterising them
is limited by  selection bias. Traditionally, clusters
have been found through optical galaxy counts, X-ray emission or
the thermal Sunyaev-Zel'dovich (SZ) effect, resulting in 
 several thousand detections. 
 
The radio emission from clusters consists of (a) the halos, which are large (up to $\sim 1$ Mpc) diffuse regions of synchrotron emission found in
galaxy cluster centres, (b) radio relics, which are diffuse areas of shock-excited synchrotron emission, (c) mini halos, found around the central AGN in some
cluster cores, and (d) tailed galaxies, which are FRI sources whose jets and lobes are blown by ram pressure from the intra-cluster 
medium.  
Diffuse emission from halos and relics has so far been identified in only a few tens of clusters.

 The radio halo power scales roughly as the
third power of cluster mass \citep[e.g.][]{Sommer14}
so current surveys have been
able to detect only the few most massive objects at relatively low
redshifts. Radio halo luminosity is also strongly bimodal \citep{Cassano12} with radio haloes being detected almost exclusively in disturbed  or merging clusters. The SKA's short baselines make SASS1 very sensitive to diffuse emission.
SASS1 will therefore be able
to detect even a $2\times10^{14}$ solar mass cluster at $z=0.5$, where its total
flux density is $\sim$ 30 \ujybm, and will detect about 100,000
radio halos out to redshift $z=0.5$. This mass limit will be comparable to numbers expected from the future X-ray ({\it eROSITA}) and optical ({\it Euclid})
missions.  Such unprecedented numbers of radio selected clusters will
 enable the measurement of  the dark matter halo merger rate with redshift, a
key but untested component of the hierarchical structure formation model. 

Other clusters will be detected from their peripheral relics or tailed radio galaxies. Based on ATLAS data, \cite{Dehghan14} suggest SASS1 will detect $\sim$1 million tailed galaxies, perhaps making them the most widespread cluster diagnostic 
in the SASS1 era. Similarly, the number of high-redshift ($z>1$) clusters is
expected to grow to tens of thousands, Cross-identifying
these objects with other probes (e.g. {\it eROSITA}) will provide a precise
constraint on the dark energy equation of state.

Clusters of galaxies are not isolated regions, but are located at the
intersection of filaments and sheets in the large-scale
structure (LSS, or the ``cosmic web''). The filaments themselves are expected to radiate radio synchrotron emission, powered by the infall shocks of
baryons.
The sensitivity and field-of-view of SASS1
will enable it to image this diffuse
 emission, thus mapping  the cosmic web. An all sky survey will accurately
map the energy distribution of the relativistic electrons in the warm hot
intergalactic material (WHIM), and cross-correlate it with other LSS
probes. 

\subsection{Evolution of Galaxies}

Extragalactic radio astronomy  tackles the  origin, assembly, and evolution of galaxies, including regulation by feedback mechanisms, and the origin, growth, and evolution of supermassive black holes.  Largely unstudied populations of starburst galaxies and low-power active galactic nuclei (AGN) are also being explored

Source counts at flux densities above 1 mJy/beam are dominated by radio-loud AGN, while below 1 mJy/beam they include significant contributions from radio-quiet AGN and star-forming galaxies.  SASS1 will not only detect the underlying populations and dissect the lifecycles of AGN \citep{Hopkins03}, but  will also answer some of most debated questions in modern astrophysics such as the importance of AGN feedback in galaxy formation and evolution across cosmic time \citep{Best12}. Radio-loud AGN also provide an obscuration-independent method of selecting the highest redshift AGN, which trace proto-clusters in the early Universe \citep{Miley08}. 

Some classes of radio AGN are rare, requiring an all-sky survey to obtain statistically significant sample sizes  across a range of luminosities and redshifts. For example, CSS and GPS sources \citep{Fanti90, Randall} represent an early stage of radio-loud AGN that are completely embedded in their host galaxies, and hence particularly important for investigations of the AGN feedback and AGN lifecycles \citep{Kap}. Even rarer are the radio-loud AGN in the brief phase as they cease jet activity, and their luminosity rapidly drops. The expected number density of dying radio sources is hard to estimate as few are known \citep{Parma07, Dwara09} and a high survey sensitivity to diffuse emission is required. Rarer still, at least in the local Universe, are the sources in which an FRII-luminosity jet has turned on inside a starburst galaxy, but has not yet bored its way through the dense molecular gas and dust to quench the star formation: only one has so far been found \citep{Mao14}. Detection of a representative number of such rare radio sources will determine  AGN duty cycles and large scale (cluster scale) feedback processes; only all-sky surveys will provide enough data to compile such samples. 

 
\subsection{The Magnetic Sky}
\label{magnet}
Magnetic fields are  important at all scales in the Universe, from  small scales in the interstellar medium, to  the vast scale of cosmic filaments.
Yet, at all scales, we fail to understand the effect and magnitude of the magnetic contribution to the energy balance and evolution. 
High quality polarisation observations
address diverse science goals,  including the effect on the formation and evolution of galaxies and clusters,   the magnitude of fields in the intra-cluster medium and their role in the cosmic web, and the question of the origin of cosmic magnetism.
The key tool for analysing polarisation data from
broadband  radio surveys is Rotation Measure (RM) Synthesis \citep{Brentjens05}, which can
detect  faint polarised emission. 

SASS1 will produce a detailed total intensity and polarization image of the entire sky, delivering an all-sky RM grid   300-1000 times  denser than those currently available  \citep{Taylor09}, and 3-10 times  denser than those produced in the ASKAP-POSSUM survey \citep{Gaensler10}, and probing 10-100 times deeper than current polarisation surveys (see Fig. 5).  RM Synthesis  of SASS1 data will recover the polarised signal affected by internal depolarisation. These data will  track the evolution of magnetic fields in the interstellar medium, in Active Galactic Nuclei, in the intracluster medium, at the boundary of galaxy clusters, in the bridges that join clusters, and in the filamentary cosmic web. There are even indications that the evolution of magnetic fields in normal galaxies at high redshifts can be traced by their Faraday effects when they are seen as MgII absorbers in front of polarized background quasars \citep{Farnes}.  

A significant step forward
would be the measurement of the redshift evolution of the Faraday
rotation measure in clusters, as SASS1 will  detect $\sim 10^{5}$
clusters with at least one background source in each \citep{Krause09}.
 Compared to the few RMs known for a few
nearby clusters today, this will revolutionize our understanding of
the cosmic magnetic fields and their impact on the   growth of large-scale structure.

\begin{figure}
\begin{center}
\includegraphics[width=8cm, angle=0]{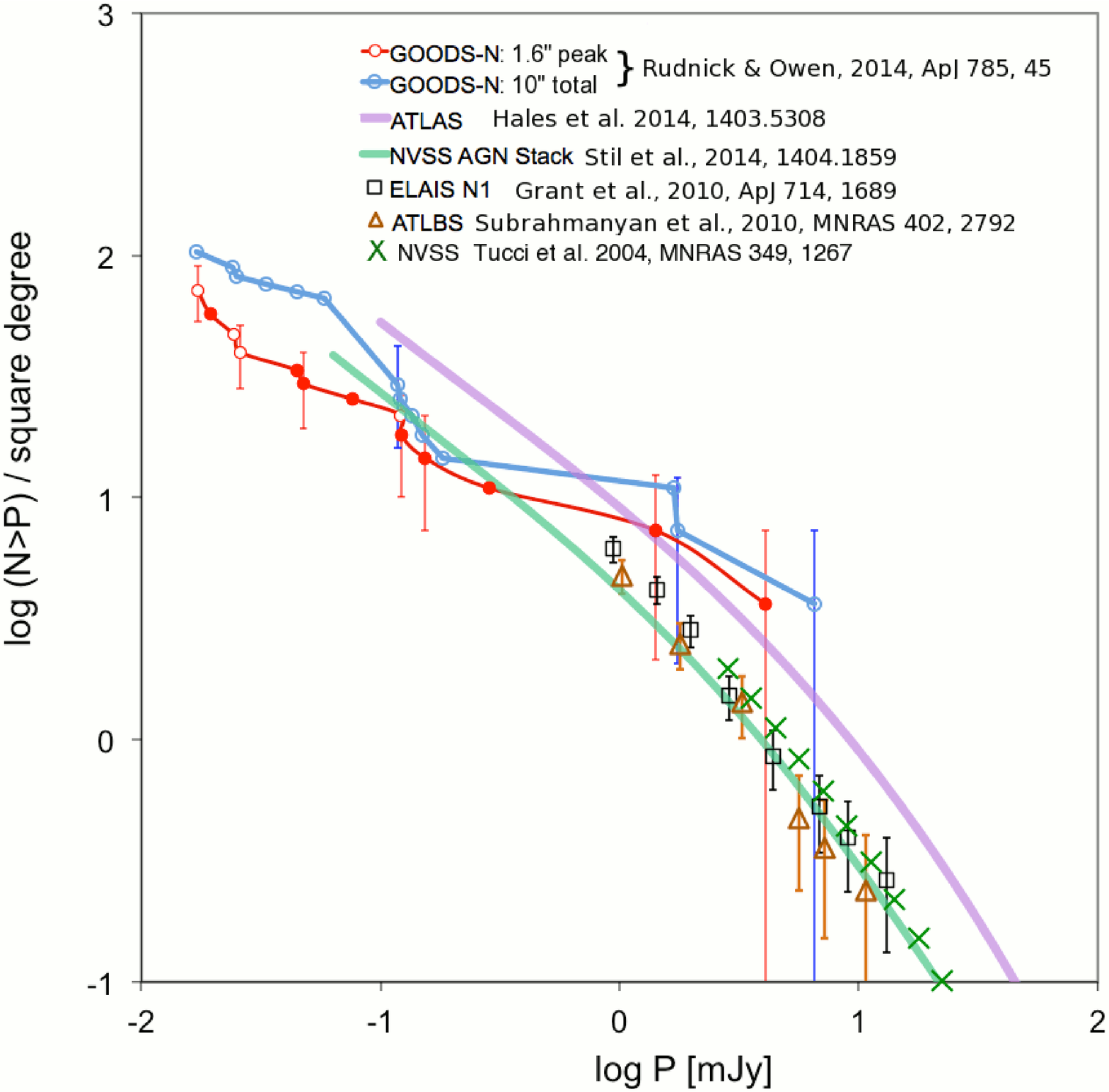}
\caption{Recent determinations of polarised number counts, taken from 
\citep{Rudnick, Hales14, Stil14, Grant10, Ravi10, Tucci04}. 
P is the total linearly polarised intensity. SKA will probe 10-100 times deeper than these surveys. }
\label{polcounts}
\end{center}
\end{figure}

\subsection{Nearby Galaxies}

SASS1 will measure star formation rates and AGN activity in thousands ($\sim$50000 galaxies at z$<$0.01) of nearby galaxies, each of which will have generous multiwavelength data. Hundreds of galaxies will have ground-based drift-scan/IFU spectroscopy, along with {\it WISE} mid-infrared images, enabling detailed spatial comparison of the radio star formation rate  with H-alpha star formation rate, with dust extinction correction. The 2-arcsec resolution of SASS1 corresponds to 400 pc at z=0.01, so that individual giant molecular clouds can be studied. 
 
Virtually all spiral galaxies and high mass elliptical galaxies in this redshift range will be detected by SASS1 \citep{Brown11}, so the radio duty cycle of galaxies can be directly measured. The shape of the radio luminosity function can thus be derived from the galaxy mass function and radio duty cycle, rather than being empirically modelled with broken power-laws. 

Since the large-scale ($\sim$10 Mpc) clustering of dark matter halos is a function of their mass, the halo masses of radio source populations will be determined by measuring how optically selected galaxies 
cluster around radio sources. At fixed halo mass, the small-scale clustering of galaxies as a function of radio power will reveal the relative contribution of galaxy mergers and secular evolution in driving star formation and AGN activity. If secular evolution plays a dominant role in driving star formation rates in nearby galaxies, the $\leq$ 1 Mpc environments of galaxies (at fixed halo mass) with high and low star formation rates will be virtually identical.

\section{Science outcomes in SKA1}
\S\S 2-4 were based on the baseline design specifications of SKA1, giving a continuum survey (SASS1) of 3$\pi$ steradians (i.e. covering the declination range -90 to +30$\deg$) with an rms sensitivity of 2 \ujybm. Table \ref{surveytable} shows that SASS1 increases the number of known radio sources by nearly an order of magnitude compared to EMU, or by a factor of 200 compared to the number ($\sim$ 2.5 million) of radio sources known in 2014, resulting in the following science outcomes:
\vspace{-3mm}
\begin{itemize}
\itemsep-5pt
\item Discovering the unexpected: experience with previous major instruments (e.g. {\it HST}) shows that the most significant discoveries from SKA will not be those listed in the science goals. SASS1 goes nearly an order of magnitude deeper than any other all-sky survey with all-sky coverage ensuring detection of  even the rarest  phenomena within that  phase space.

\item Transforming radio-astronomy from niche to mainstream: SASS1 sensitivity will detect most star-forming galaxies to $R\sim$23, matching SDSS and SkyMapper, so SASS1 will supply radio data for most galaxies currently being studied by optical astronomers, and will be routinely used in constructing SEDs and photo-z's.
\item Cosmology: SASS1 will place significant constraints on the parameters of dark energy, modified gravity, non-gaussianity, and neutrino mass, and will test low multipole isotropy of the Universe, giving  measurements independent of those  from optical and HI surveys.
\item Galaxy Clusters and Large-scale Structure: SASS1 will detect about a million clusters, including about 100,000 radio halos. Together with {\it Euclid} and {\it eROSITA} data, this will transform our understanding of the physics of the large scale structure of the Universe.
\item Evolution of Galaxies: SASS1 will detect about 500 million galaxies spanning all redshifts, and will trace the growth of black holes, the evolution of the cosmic star formation rate, and the interaction between these, to exquisite precision, finally nailing down the feedback mechanisms which regulate the growth and evolution of galaxies.
\item The Magnetic Sky: SASS1 will measure the polarisation of about 10 million galaxies, determining not only the effect of magnetic fields on galaxies, but also providing a rotation measure grid nearly three orders of magnitude denser than currently available, measuring the intergalactic magnetic field as a function of redshift, giving clues to the origin of magnetism.
\item Nearby Galaxies: SASS1 will provide detailed imaging of virtually all nearby (z$<$0.01) star-forming and high-mass elliptical galaxies, to a resolution of 400pc or higher, measuring the effect of both  environment and AGN activity on the evolution of these galaxies.
\end{itemize}

\section{Science outcomes from SKA1 early science operations}
Here we estimate the scientific productivity of an ``early science SKA1'' all-sky-survey, with 50\% of the sensitivity of SKA2. We assume 
the ``50\%'' SKA1 still has the same resolution as the full SKA1, and is therefore able to deliver an-all-sky radio continuum survey with 4 \ujybm\ rms, which is  a factor 2.5 deeper than the deepest previous survey, ASKAP-EMU.
\begin{itemize}
\itemsep-5pt
\item Discovering the unexpected: Halving SKA1 sensitivity makes it only a factor of 2.5 deeper than EMU, reducing the amount of virgin observational space, reducing the likelihood of new discoveries compared to SKA1.
\item Transforming radio-astronomy from niche to mainstream: Halving SASS1 sensitivity will detect all star-forming galaxies to $R\sim$22 so will still supply radio data for most galaxies being studied optically, and will be increasingly used in constructing SEDs and photo-z's.
\item Cosmology: A ``half-SASS1'' will still make important  cosmological measurements, but they will be less competitive than those derived from optical or HI measurements.
\item Galaxy Clusters and Large-scale Structure: Halving the sensitivity of SASS1 will result in about one quarter as many clusters detected, but this is still an enormous increase on what is currently available at present, or will be available from EMU. 
\item Evolution of Galaxies: Halving the sensitivity of SASS1 significantly reduces the redshift range at which star-forming galaxies and low power AGN can be detected. The study of high-power AGN will be largely unaffected, but measuring the evolution of the cosmic star formation rate will be seriously affected.
\item The Magnetic Sky: Halving the sensitivity of SASS1 will result in a reduction by one third in the number of galaxies for which polarisation can be measured. The increase compared to ASKAP-POSSUM is still valuable as the measurement of the  intergalactic magnetic field depends critically on the sampling density.
\item Nearby Galaxies: Halving SASS1 sensitivity reduces the number of nearby galaxies by a factor of $\sim$3,  leaving $\sim$20,000 which can be studied in detail, far more than  available now.
\end{itemize}

\section{Science outcomes from SKA2}
Assuming that SKA2 is ten times more sensitive and twenty times  the FOV of SKA1-MID, then it can survey the whole sky in 83 12-hour observations. A 6-month survey (SASS2) can therefore observe each region of the sky four times, yielding an rms sensitivity 20 times deeper than SASS1.  In 6 months, SASS2 will have rms $\sim$0.1\ujybm\ at $< $0.1arcsec (assuming the longer baselines planned for SKA2). The higher resolution will also provide low confusion and unambiguous cross-identifications to optical and infrared counterparts. Table \ref{surveytable} shows that SASS2 increases the number of known radio sources by nearly three orders of magnitude compared to current knowledge. This results in the following science outcomes:
\vspace{-3mm}
\begin{itemize}
\itemsep-5pt
\item Discovering the unexpected: Figure 1 shows that SASS2 covers orders of magnitude more observational phase space than any other radio survey, almost certainly resulting in major science discoveries that are unlikely to  feature in the current SKA science goals.
\item Transforming radio-astronomy from niche to mainstream: SASS2  will detect all star-forming galaxies to $R\sim$27, matching {\it LSST}, and will supply radio data for nearly all galaxies  studied by optical astronomers, becoming an indispensable component of SEDs and photo-z's.
\item Cosmology: SASS2 will probably measure the parameters of dark energy, modified gravity, non-gaussianity, and neutrino mass with lower uncertainties than other competing projects ({\it DES, Euclid}, etc) but this has yet to be established by careful modelling. Figure 5 gives on example of how SASS2 can measure non-gaussianity ten times better than {\it Euclid}, provided SASS2 sources can be placed into 3 redshift bins.
\item Galaxy Clusters and Large-scale Structure: SASS2 could, in principle, detect about 30 million clusters from their radio emission, including about 2 million radio halos. However, this figure is extrapolated far beyond our current knowledge of cluster physics, and  is unlikely to be accurate. Clearly, SASS2 will be venturing into uncharted territory in this field!
\item Evolution of Galaxies: SASS2 will detect over 3 billion galaxies spanning all redshifts. It will trace the growth of black holes, the evolution of the cosmic star formation rate, and the interaction between these to high precision, and will discover new classes of galaxy. SASS2's high resolution will enable unambiguous identification with optical and IR sources.
\item The Magnetic Sky: SASS2 will measure polarisation for about 70 million galaxies, measuring not only  the effect of magnetic fields upon galaxies, but also providing a rotation measure grid nearly four orders of magnitude denser than currently available. It will measure the intergalactic magnetic field as a function of redshift and environment (sheet, string, void, etc.) giving vital clues to the origin and evolution of cosmic magnetism.  
\item Nearby Galaxies: SASS2 will not only provide detailed imaging of all nearby (z$<$0.05) star-forming and high-mass elliptical galaxies, but will do so at a resolution of   0.1 arcsec (100pc or higher), measuring the effect of both  environment and AGN activity on the evolution and star formation in these galaxies.
\end{itemize}

{\bf Acknowledgment:} We thank an anonymous referee for careful reading and helpful comments. ADK acknowledges financial support from the Australian Research Council Centre of Excellence for All-sky Astrophysics (CAASTRO), through project number CE110001020.

\end{document}